\documentclass{INTERSPEECH2023}

\interspeechcameraready 

\usepackage{booktabs}
\usepackage{multirow}
\usepackage{array}
\usepackage{float}

\usepackage{color}
\usepackage{xcolor}
\usepackage{kotex}

\newcommand\sh[1]{\textcolor{black} {#1}}
\newcommand\bg[1]{\textcolor{black} {#1}}
\newcommand\sm[1]{\textcolor{black} {#1}}

\title{Improving Small Footprint Few-shot Keyword Spotting with Supervision on Auxiliary Data}
\name{Seunghan Yang, Byeonggeun Kim, Kyuhong Shim, Simyung Chang}
\address{Qualcomm AI Research${}^{\dagger}$, Qualcomm Korea YH, Seoul, Republic of Korea\thanks{${}^{\dagger}$Qualcomm AI Research is an initiative of Qualcomm Technologies, Inc.}}
\email{\{seunghan, kshim, simychan\}@qti.qualcomm.com}

\begin{document}

\maketitle
 
\begin{abstract}
Few-shot keyword spotting (FS-KWS) models usually require large-scale annotated datasets to generalize to unseen target keywords. However, existing KWS datasets are limited in scale and gathering keyword-like labeled data is costly undertaking. To mitigate this issue, we propose a framework that uses easily collectible, unlabeled reading speech data as an auxiliary source. Self-supervised learning has been widely adopted for learning representations from unlabeled data; however, it is known to be suitable for large models with enough capacity and is not practical for training a small footprint FS-KWS model. Instead, we automatically annotate and filter the data to construct a keyword-like dataset, LibriWord, enabling supervision on auxiliary data. We then adopt multi-task learning that helps the model to enhance the representation power from out-of-domain auxiliary data. Our method notably improves the performance over competitive methods in the FS-KWS benchmark.
\end{abstract}

\section{Introduction}
{
Few-shot keyword spotting (FS-KWS) refers to the task of recognizing specific keywords in an audio signal, where only a limited number of examples are available for each keyword. This task has gained significant attention in recent years due to its practical importance in various applications such as voice assistants requiring user-defined keywords~\cite{kim2019query, FSKWS_loss1, FSKWS_loss2, FSKWS_loss3, learning_3, learning_2, learning_1}.
To solve this task, few-shot learning approaches have been proposed to learn a discriminative embedding space that can generate general representations for novel classes only given few examples.
They mainly leverage prior knowledge from related tasks or similar data distributions using large-scale annotated training datasets~\cite{protonets, finn2017model, largescale_fewshot_1, largescale_fewshot_3}.
}

\begin{figure}[t]
  \centering
  \includegraphics[width=0.84\linewidth]{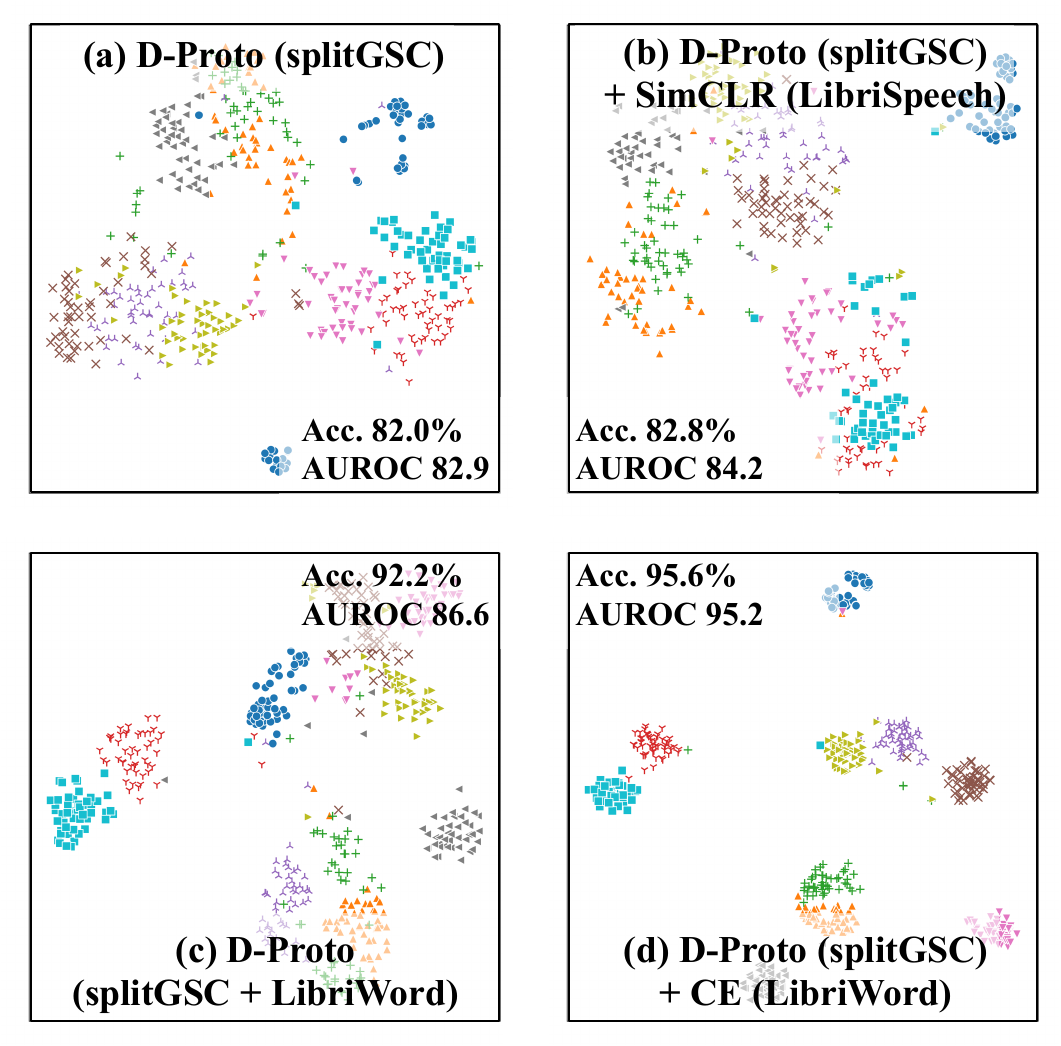}
\vspace{-3.5mm}
  \caption{
\textbf{\sm{t-SNE Visualization of Unseen Keywords Embeddings Using Various Learning Methods on the FS-KWS Benchmark~\cite{FSOSR}.}}
We train four models using in-domain splitGSC and out-of-domain reading speech datasets, including LibriSpeech and our proposed word-level labeled dataset, LibriWord: \textbf{(a)} metric learning~\cite{FSOSR} on splitGSC, \textbf{(b)} (a) + self-supervised learning~\cite{SimCLR} on LibriSpeech, \textbf{(c)} metric learning on both datasets (splitGSC + LibriWord), and 
\sm{\textbf{(d)} our proposed AuxSL on splitGSC and LibriWord.}
The closed-set classification accuracy (Acc.) and open-set detection rate (AUROC) indicate the performance of the models.
}  
  \label{fig1}
\vspace{-7mm}
\end{figure}

However, the keyword spotting (KWS) datasets~\cite{GSC, Heysnips, MCKWS} that are accessible to the public are typically limited in size and have a small number of categories. Acquiring a comprehensive annotated dataset suitable for KWS is a costly work.
Given the existing datasets, the performance of FS-KWS is limited, as reported in Fig~\ref{fig1}(a), even with the use of an advanced few-shot learning technique~\cite{FSOSR}.
To this end, we propose to exploit auxiliary data, especially reading speech data from audiobooks, such as LibriSpeech~\cite{librispeech} and MLS~\cite{pratap2020mls}, which are readily publicly available in large quantities and easily collectable.
These data are not built for the keyword spotting task ({\it e.g.}, LibriSpeech~\cite{librispeech} for speech recognition), but they are expected to help the model learn a more robust embedding space.

One way to leverage knowledge from a large-scale unlabeled dataset is through self-supervised learning (SSL)~\cite{SimCLR, simsiam, wav2vec, Hubert}.
SSL has gained widespread usage in various applications due to its many advantages, particularly its ability to produce robust feature representations without the need for labeled data.
SSL has been successfully employed in the pre-training stage of the model~\cite{simsiam, Hubert} or in combination with a target objective for fine-tuning the model~\cite{s4l, cui2022exploring}.
However, prior literature has established that SSL is only beneficial to large models, as evidenced by~\cite{simclr2, SEED}.
Therefore, the lightweight nature of FS-KWS models presents a challenge for the use of SSL.
In this paper, we empirically derive that the existing SSL methods are also not suitable for training lightweight models in FS-KWS, as shown in Fig.~\ref{fig1}(b).
Instead, we automatically annotate the reading speech data with a word extraction technique~\cite{mcauliffe2017montreal} and construct a well-organized keyword-like dataset called \bg{\textit{LibriWord}} by filtering and balancing the data, effectively enabling the use of auxiliary data with supervision on lightweight models.

\sm{To effectively leverage both in-domain command-like data and out-of-domain auxiliary data, we propose a simple but effective multi-task learning (MTL) framework, called AuxSL, that incorporates an additional classifier with a supervised loss function for the out-of-domain data.}
\sm{AuxSL} enables the model to enhance the representation power from out-of-domain auxiliary data while primarily learning keyword representations on in-domain data. The naive approach of combining the two datasets and applying a single loss function to the entire dataset, as shown in Fig.~\ref{fig1}(c), results in limited performance improvement.
Our proposed framework is evaluated through extensive experiments on the FS-KWS benchmark, demonstrating its superior discriminative capabilities. In Fig.~\ref{fig1}(d), the results show a relative improvement of $16\%$ in both closed- and open-set settings when using 5-shot samples, compared to the baseline.

\section{Related Works}
\subsection{Small Footprint Keyword Spotting}
The aim of conventional keyword spotting is to detect a small set of pre-defined speech signals, such as the activation phrases ``Alexa" and ``OK Google". Recent research has primarily focused on enhancing accuracy while minimizing memory requirements and reducing power consumption, allowing for the deployment to an always-on system~\cite{bcresnet,  dsresnet, lopez2021novel, yang2022personalized}. Some studies have centered around the development of optimal architectures for keyword spotting~\cite{bcresnet,  dsresnet}, while others have aimed to improve the loss function~\cite{lopez2021novel} or training approach~\cite{yang2022personalized}.
\subsection{Few-shot Keyword Spotting}
Few-shot keyword spotting (FS-KWS) aims to support systems that require user-defined keywords. Unlike conventional keyword spotting, FS-KWS involves a scenario in which new keywords that were not seen during training are enrolled and tested on the device.
To achieve FS-KWS, it is crucial to learn robust representations in the pre-training stage using a large-scale dataset. To further improve performance, various loss functions~\cite{FSKWS_loss1, FSKWS_loss2, FSKWS_loss3}, architectures~\cite{hou2021meta}, and learning strategies~\cite{learning_3, learning_2, learning_1} have been proposed.
Modifying the learning strategies is closely related to our work. Kao et al.~\cite{learning_3} present a two-stage framework that compares several popular SSL models to determine the best model for FS-KWS. Lee et al.~\cite{learning_2} incorporate auxiliary synthetic data generated by TTS, while Shin et al.~\cite{learning_1} propose a novel approach that leverages the linguistically corresponding patterns between speech and text sequences.
Our approach is distinct from prior studies as we propose utilizing auxiliary data from reading speech domain along with a supervised multi-task learning strategy, with the objective of training lightweight models. Recently, D-ProtoNets~\cite{FSOSR} has been proposed for few-shot open-set keyword spotting, which involves open-set classes during test time. Our proposed method is expected to improve open-set performance in addition to closed-set performance, as it helps to create robust representations.

\section{Proposed Methods}
\subsection{Problem Definition}
{\bf Training objectives for few-shot learning:}
In the domain of few-shot learning, metric learning based techniques have been predominantly investigated for constructing a robust embedding space that enables models to extract discriminative embeddings for unseen keywords and perform keyword detection using distance metrics.
One of the most popular metric learning-based methods is ProtoNets~\cite{protonets}.
At each training iteration, ProtoNets create a classification scenario by constructing a $N$-way and $K$-shot problem from training data, where $N$ represents the number of classes and $K$ denotes the number of support samples per class.
ProtoNets first construct the prototype of each class by averaging the embeddings of support samples, $c_{n} = \frac{1}{K}\sum_{i=1}^{K}F_{\theta}(x^{s}_{n,i})$, where $F_{\theta}(\cdot)$ is a feature extractor.
Subsequently, the prototypical loss function forces the query samples to minimize the distance from the corresponding prototypes as follows:
\vspace{-1mm}
\begin{equation*}
L_{FSL} = -\frac{1}{M}\sum_{i=1}^{M}\text{log }p_{\theta}(y=n|x^{q}_{n,i}), \quad \text{where}
\end{equation*}
\begin{equation}
\label{eq:cls_prob}
p_{\theta}(y=n|x^{q}_{n,i}) = \frac{\text{exp}(-d(F_{\theta}(x^{q}_{n,i}), c_{n}))}{\sum_{n'=1}^{N}\text{exp}(-d(F_{\theta}(x^{q}_{n,i}), c_{n'}))}.
\end{equation}
$d(\cdot)$ can be any distance metric, $x^{q}_{n,i}$ is $i$-th query sample of class $n$, and $M$ indicates the number of query samples of class $n$ in the $N$-way and $K$-shot problem.
At the inference stage, prototypes of new classes are constructed by averaging the embeddings of enrolled samples. The test samples are then classified based on the \bg{classification probability in Eq.~\ref{eq:cls_prob}.}

{\bf \noindent Limitation of few-shot keyword spotting:}
Despite the success of few-shot learning, a large-scale dataset is required to learn a robust embedding space.
For keyword spotting, the dataset\bg{s usually have limited sizes,} making it difficult to learn general keyword representations. While collecting in-domain command-like data is an effective way to create a robust embedding space, it is challenging to obtain such datasets. To address this issue, we propose to use easily obtainable auxiliary data, especially reading speech data.

\begin{table}[]
\centering
  \caption{Metadata of LibriSpeech and proposed LibriWord.}
  \vspace{-3mm}
\resizebox{0.75\linewidth}{!}{
\begin{tabular}{ccc}
\toprule
\multicolumn{1}{c|}{Info.}          &              \multicolumn{1}{c|}{LibriSpeech} & LibriWord   \\ \midrule
\multicolumn{1}{c|}{\# of samples}         &              \multicolumn{1}{c|}{4,806,675}                       & 300,000                                                                             \\
\multicolumn{1}{c|}{\# of keywords}        &              \multicolumn{1}{c|}{58,997}                 & 1,000                                                                               \\
\multicolumn{1}{c|}{Audio length}          &              \multicolumn{1}{c|}{[0.0s, 11.1s]}                 & {[0.3s, 2.2s]}                                                                    \\
\multicolumn{1}{c|}{Annotations}           &              \multicolumn{2}{c}{Word-level labels}                                                                  \\
\multicolumn{1}{c|}{Annotator}             &              \multicolumn{2}{c}{Montreal Forced Aligner~\cite{mcauliffe2017montreal, lugosch2019speech}}               \\ \bottomrule
\end{tabular}}
\label{data_metadata}
\vspace{-5mm}
\end{table}

\subsection{LibriWord}
Instead of collecting human-labeled keyword spotting data, we create a dataset named LibriWord containing segmented utterances and corresponding word-level labels. The samples are obtained from the LibriSpeech corpus~\cite{librispeech}, which comprises roughly 1,000 hours of read English speech at 16kHz. It contains numerous words, but lacks word-level alignments and only utterance-level transcriptions are provided. To obtain word-level segmented samples, we employ the Montreal Forced Aligner~\cite{lugosch2019speech, mcauliffe2017montreal}, a word extraction technique previously used in~\cite{learning_1, lugosch2018donut}. We then form a balanced dataset by organizing the extracted words based on the number of samples and eliminating similar keywords.
Specifically, when a word partially overlapped with another word, such as with past tense, plural forms, or negative forms, we randomly remain one of them to simplify the representation learning process for small models.
As a result, LibriWord includes 300 samples for each top 1,000 frequently appearing keywords. Table~\ref{data_metadata} presents the metadata before and after the refinement process, respectively. 
Learning with LibriWord, a comparatively smaller dataset than LibriSpeech, yields the benefits of reducing the burden of collecting auxiliary data and saving storage space, while achieving better performance (please see Section \ref{libriword_ablation}).

\subsection{\sm{FS-KWS with Auxiliary Supervision on LibriWord}}
\begin{figure}[t]
  \centering
  \includegraphics[width=1.0\linewidth]{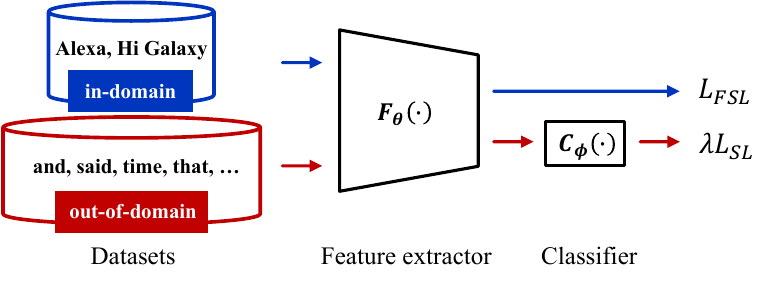}
 \vspace{-7mm}
  \caption{
  \sm{\textbf{Our Proposed AuxSL Framework.}}
$F_{\theta}(\cdot)$ is trained with both in-domain data for keyword spotting and out-of-domain auxiliary data from the reading speech source, and $C_{\phi}(\cdot)$ is trained to classify auxiliary words for helping to learn robust representations.
  }
\vspace{-6.5mm}
  \label{fig2}
\end{figure}

\sm{
We propose a simple but effective multi-task learning framework, called AuxSL, which utilizes an additional classifier and supervised loss function for the out-of-domain auxiliary data to mitigate potential representation skew that may arise when directly sharing the feature extractor between in-domain and out-of-domain data, motivated by~\cite{zhou2018deep}. Our multi-task learning objectives are calculated as follows:
}
\vspace{-1mm}
\begin{equation}
\vspace{-.5mm}
L_{AuxSL} = L_{FSL} + \lambda L_{SL},
\end{equation}
where $L_{FSL}$ represents any few-shot learning loss function on in-domain training data and $L_{SL}$ represents supervised loss function on out-of-domain auxiliary data.
$\lambda$ is a balancing parameter for auxiliary loss.
Here, we maintain metric learning loss on in-domain data and use conventional cross-entropy loss, as depicted in Figure~\ref{fig2}.
This approach does not incur additional cost during inference since the classifier is not used.

For our experiment, we employ a dummy prototypical loss function of D-ProtoNets~\cite{FSOSR} for $L_{FSL}$ to efficiently handle both closed- and open-set query samples. 
D-ProtoNets~\cite{FSOSR} are trained using the dummy prototypical loss, which incorporates a learnable open-set prototype specifically designed to represent the open-set class.
The open-set prototype is trained jointly with the class-wise prototypes to enable query samples of both closed- and open-set to be closely associated with their corresponding prototypes in the $N+1$ classification task.
During inference, if the probability of a query test sample $x^{q}_{t}$ belonging to the open-set class $N+1$, {\it i.e.}, $p_{\theta}(y=N+1|x^{q}_{t})$, exceeds a pre-defined threshold, then it is verified as the open-set class.

\section{Experiments}
\subsection{Experimental Settings}
{\bf Dataset.}
We use a standard benchmark splitGSC~\cite{FSOSR} on the Google Speech Commands (GSC) dataset~\cite{GSC} for keyword spotting task.
The splitGSC contains the train, validation, and test split designed for few-shot closed- and open-set keyword spotting. The split includes 15, 10, and 10 keywords for training, validation, and testing, respectively, with 24,400, 4,007, and 4,482 samples, respectively. Note that a silence category is included in all sets. The official background noise provided by GSC with a probability of 0.8 is used. See \cite{FSOSR} for more details.

{\noindent \bf Implementation details.}
We use three different small footprint backbone models, BC-ResNet8~\cite{bcresnet} and Res12~\cite{res12}, both of which take 40-dimensional log Mel spectrograms as input with a window length of 30 ms and frame shift of 10 ms, and DS-ResNet18~\cite{dsresnet}, which takes 40-dimensional Mel-frequency cepstrum coefficient features as input.
The network size of BC-ResNet8, Res12, and DS-ResNet18 is 321k, 8M, and 72k, respectively.
Each model is trained for 100 epochs using the Adam optimizer with an initial learning rate of 0.001, which is step decayed by a factor of 0.5 every 20 epochs.
Each epoch consists of 500 episodes, each containing 5 closed- and 5 open-set classes, with 5 support samples for prototypes and 5 query samples for each class.
For multi-task learning, we use a batch size of 64 with parallel sampling as in the episodic iterations, and set $\lambda$ to 1.0. As an additional module, we used a 1-FC layer as the classifier.
To evaluate the trained model, we used $1,000$ episodes at test time, with 1 or 5 support samples and 15 query samples per class including open-set classes.
We reported the average 1-shot and 5-shot accuracy and threshold-free area under the receiver-operating characteristics (AUROC) for closed- and open-set performance with 3 different seeds.

\begin{figure}[t]
  \centering
  \hspace{-2mm}\includegraphics[width=0.9\linewidth]{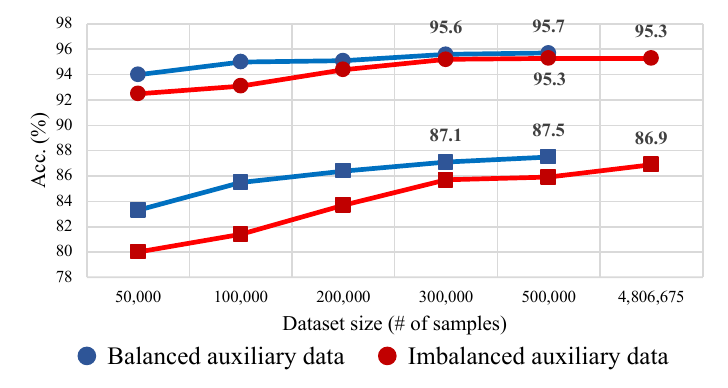}
  \vspace{-4mm}
  \caption{The accuracy of models trained with balanced and imbalanced auxiliary data on BC-ResNet8 using the MTL path (our optimal framework). The 5-shot and 1-shot performance are denoted by circle and square, respectively.}
  \label{fig3}
  \vspace{-6.5mm}
\end{figure}

{\setlength{\tabcolsep}{1.4em}
\begin{table*}[t]
\centering
  \caption{A performance comparison between models trained using self-supervised learning and supervised learning on auxiliary data.
  \sh{The Method column indicates the model training method, and the used dataset is indicated in parentheses.}
  PreT-Big uses the pre-trained \sm{large} feature extractor on Librispeech provided by torchaudio. In contrast, other methods use BC-ResNet8 as the feature extractor. We show 5-shot and 1-shot closed- and open-set average performance and standard deviation.}
  \vspace{-3mm}
\resizebox{1.\linewidth}{!}{
\begin{tabular}{c|c|c|cc|cc}
\toprule
\multirow{2}{*}{Learning scheme on auxiliary data} & \multirow{2}{*}{Training strategy} & \multirow{2}{*}{Method} & \multicolumn{2}{c|}{5-shot} & \multicolumn{2}{c}{1-shot} \\
                                 &                                  &                                & Acc. (\%)      & AUROC     & Acc. (\%)      & AUROC     \\ \midrule
                     X            &   {Baseline (321k)}            & {D-Proto (splitGSC)}                       &  82.0 (0.7)   & 82.9 (0.1) & 65.8 (0.3)    & 75.7 (0.1) \\  \cmidrule{1-7}
\multirow{9}{*}{Self-supervised learning} & \multirow{2}{*}{PreT-Big (300M)}&   Wav2vec (LibriSpeech) $\rightarrow$ D-Proto (splitGSC)  & 85.3 (0.9)     & 76.7 (1.2)  &  66.8 (1.5)    &  70.5 (0.4)  \\
                                &                                   & Hubert (LibriSpeech) $\rightarrow$ D-Proto (splitGSC)   & 90.4 (0.3) & 85.2 (0.6) &  76.2 (0.8)  &  77.3 (0.9) \\ \cmidrule{2-7}
                                & \multirow{2}{*}{PreT}              & SimCLR (LibriSpeech) $\rightarrow$ D-Proto (splitGSC)   & 79.9 (1.0)  &  81.2 (1.0)   &   64.7 (1.0)  &  75.0 (0.8)  \\ 
                                &                                    & BYOL (LibriSpeech) $\rightarrow$ D-Proto (splitGSC)   & 82.0 (1.5) & 82.7 (1.0) & 65.7 (1.6) & 76.2 (1.0) \\ \cmidrule{2-7}
                                & \multirow{4}{*}{MTL}               & SimCLR (LibriSpeech) + D-Proto (splitGSC)  & 83.0 (0.8)   &  83.1 (0.9)  &  66.6 (1.6)  &  76.0 (1.0) \\ 
                                &                                    & BYOL (LibriSpeech) + D-Proto (splitGSC)  &  81.7 (0.0)  &  82.7 (0.0)  &  65.2 (0.4)  &  76.0 (0.5) \\ \cmidrule{3-7}
                                &                                    & KD w/ Wav2vec (splitGSC) + D-Proto (splitGSC) & 82.4 (0.7)   & 83.3 (0.1)  &  66.4 (0.6)     &  76.4 (0.2)  \\ 
                                &                                    & KD w/ Hubert (splitGSC)  + D-Proto (splitGSC) & 83.7 (0.5)  & 84.4 (0.5)  & 67.8 (0.6)      &  77.3 (0.9)  \\ \midrule
\multirow{4}{*}{\bf{Supervised learning}} & PreT                     & CE (LibriWord) $\rightarrow$ D-Proto (splitGSC)    &  82.8 (1.6)   & 84.2 (1.1)  &  68.7 (1.5)    &   77.3 (0.8) \\ \cmidrule{2-7}
                                & ALL                      & D-Proto (LibriWord + splitGSC)   &  92.2 (0.9)   & 86.6 (1.3)  &  77.4 (2.1)    &   75.3 (1.6) \\ \cmidrule{2-7}
                                & \multirow{1}{*}{\bf{\sm{Ours: AuxSL}}}          & \bf{CE (LibriWord) + D-Proto (splitGSC)}        &  \bf{95.6 (0.5)}  &  \bf{95.2 (0.4)}  &  \bf{87.1 (0.8)}  &  \bf{89.4 (0.3)}  \\ \bottomrule
\end{tabular}}
\label{table2}
\vspace{-5mm}
\end{table*}
}

\subsection{\sm{Analysis of LibriWord Dataset}}
\label{libriword_ablation}
In Figure~\ref{fig3}, we present the results obtained by training with different dataset constructions after extracting words from the imbalanced Librispeech corpus, which has highly skewed distribution of samples among its words.
Notably, in this experiment, we employ our final MTL architecture solely to investigate the impact of LibriWord's dataset configuration.
Our findings indicate that constructing a dataset with a balanced number of samples per keyword leads to superior performance compared to using an equally imbalanced dataset or the entire Librispeech dataset, which is roughly 16 times larger than LibriWord.
Using a balanced number of data per keyword during training aids in building robust feature embeddings, a phenomenon that has also been observed in ImageNet, where research on handling imbalanced datasets is actively being pursued~\cite{Kang2020Decoupling, ren2020balanced}.
To address the drawbacks of imbalanced datasets, we constructed LibriWord in a balanced manner and demonstrate its empirical efficacy.

\begin{table}[]
\centering
  \caption{FS-KWS performance on various backbones.}
  \vspace{-2.5mm}
\resizebox{\linewidth}{!}{
\begin{tabular}{c|c|cc|cc}
\toprule
\multirow{2}{*}{Training strategy} & \multirow{2}{*}{Backbones} & \multicolumn{2}{c|}{5-shot} & \multicolumn{2}{c}{1-shot} \\
                                   &                                & Acc. (\%)      & AUROC     & Acc. (\%)      & AUROC     \\ \midrule
\multirow{2}{*}{Baseline}         & Res12                         &  86.8 (0.2)   &   85.9 (0.3)  &   70.4 (0.2)  &  78.5 (0.1) \\
                                   & DS-ResNet18                    & 69.2 (2.9)  &  74.8 (1.7)  &  53.7 (2.2)  &  70.3 (1.2)  \\ \midrule
\multirow{2}{*}{PreT (SimCLR)}     & Res12                         &  83.3 (1.1)  &  83.8 (0.9)   &  68.7 (1.5)  &  77.1 (0.5) \\
                                   & DS-ResNet18                    &   59.0 (2.7)   & 69.3 (0.9)   &  45.8 (2.0)   &  65.7 (0.3)  \\ \midrule
{KD (Hubert)}                       & Res12                          &  87.0 (0.5)   & 85.3 (0.1)  &  70.3 (0.8)   & 78.1 (0.4) \\ 
  {KD (Wav2vec)}                   & DS-ResNet18                    &  70.8 (0.5)  &  75.5 (0.6)   &   55.0 (0.3)  &  70.6 (0.3)  \\ \midrule
\multirow{2}{*}{ALL}               & Res12                          &  93.1 (0.2)    &  86.9 (1.7) &   78.3 (0.4)  &  75.4 (1.0) \\ 
                                   & DS-ResNet18                     &  88.0 (1.3)  &  82.1 (2.2)  &  71.7 (1.2)  &  72.1 (2.3) \\\midrule
\multirow{2}{*}{\bf{\sm{Ours: AuxSL}}}    & Res12                           & \bf{95.6 (0.3)}  &  \bf{94.1 (0.6)}  &  \bf{85.5 (0.2)}  &  \bf{84.3 (1.4)} \\
                                   & DS-ResNet18                     & \bf{90.8 (0.3)}  &  \bf{89.2 (0.9)}  &  \bf{77.4 (1.0)}  &  \bf{80.2 (1.8)} \\ \bottomrule
\end{tabular}}
\label{table3}
\vspace{-6mm}
\end{table}

\subsection{\sm{Comparison of SSL and SL Methods on Auxiliary Data}}
In Table~\ref{table2}, we compared the performance of few-shot keyword spotting models trained using self-supervised learning (SSL) and supervised learning (SL) on auxiliary data. The baseline model, D-Proto~\cite{FSOSR}, was trained with SL on splitGSC without using auxiliary data. We evaluated three SSL methods: (1) PreT-Big, which uses a large-scale pre-trained feature extractor from LibriSpeech and fine-tunes only the classifier for keyword spotting, (2) PreT, which pre-trains a keyword spotting model on LibriSpeech using SimCLR and BYOL, \sh{(3) MTL with SSL, which uses SSL on LibriSpeech and SL on splitGSC, and (4) MTL with knowledge distillation (KD), which uses feature distillation~\cite{adriana2015fitnets} from the large-scale pre-trained feature extractor and SL on splitGSC.}
\sh{We also evaluated three SL methods: (1) PreT, which pre-trains the keyword spotting model using cross-entropy loss (CE),} (2) ALL, which uses all data together to train the model, and (3) \sm{AuxSL}, our proposed method that uses a different path for each dataset. Balancing parameters for MTL methods are set to 1 and all hyper-parameters are chosen based on validation performance.

{\noindent \bf Self-supervised learning on auxiliary data.}
 It is well-known that self-supervised learning is challenging for small models~\cite{simclr2, SEED}, and we empirically demonstrate that SSL is also ineffective for lightweight keyword spotting models. Despite extensive data augmentation and hyperparameter tuning, MTL with SSL and KD achieve limited performance improvement. Moreover, PreT degrades performance compared to the baseline, which trains the model on randomly initialized parameters.
In small-footprint keyword spotting models, learning invariant information with SSL-based pre-training hinders the creation of keyword representations.
Among the SSL methods, PreT-Big shows significantly better performance than the baseline. However, PreT-Big utilizes large feature extractors of size 300M, while other models use BC-ResNet8 that size is around 300K.

{\noindent \bf Supervised learning on auxiliary data.}
Different from PreT with SSL, pre-training with SL boosts the performance of the baseline, which helps to learn keyword representations.
We observe that the ALL method, which uses both datasets together for FS-KWS, is more beneficial for representation learning of small models than SSL, even in the presence of a domain gap between the two datasets. This naive approach benefits from well-organized word-like speech data for keyword representation learning.
Our AuxSL adopts a different path for the auxiliary data through the MTL framework, which results in significant performance improvement as the model learns useful information for the target task from the auxiliary data.


\vspace{-1mm}
\subsection{\sm{Evaluation on Various Architecture and Model Size}}
To evaluate the effectiveness of our proposed method on various architectures and model sizes, we conducted experiments using three different backbone architectures: BC-ResNet, Res12, and DS-ResNet18. We applied the same training and evaluation settings as in the first sub-section, except for changing the backbone architecture.
As shown in Table~\ref{table3}, we apply the competitive methods to other backbones, Res12 and DS-ResNet18, but the performance improvements are marginal, similar to the results obtained with BC-ResNet8. Our proposed MTL approach outperforms other training strategies.

\begin{figure}[t]
  \centering
  \hspace{-2mm}\includegraphics[width=1.025\linewidth]{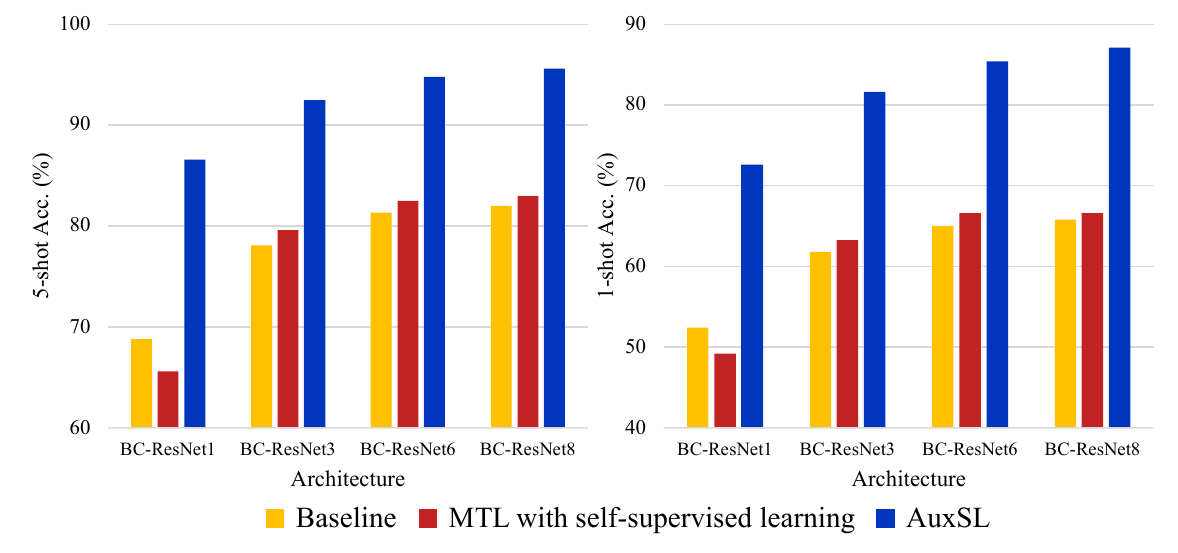}
  \vspace{-7mm}
  \caption{A performance comparison on various model size.}
  \label{fig4}
  \vspace{-7mm}  
\end{figure}

In Fig.~\ref{fig4}, we observe that the performance of all methods, including the baseline, SSL, and SL, improves as the feature extractor size increases. However, SSL decreases performance for tiny models, such as BC-ResNet1 (9.2K), compared to the baseline. Remarkably, our proposed \sm{AuxSL} approach applied to the smallest model (BC-ResNet1) outperforms other methods, including the largest model (BC-ResNet8).
This result demonstrates the effectiveness of our method in learning discriminative representations even in extremely low-resource settings.

\section{Discussion and Conclusion}
In this paper, we addressed the challenge of few-shot keyword spotting (FS-KWS) by proposing a framework that leverages commonly available reading speech data as auxiliary data. Our approach has two main contributions: (1) the creation of a well-organized and balanced keyword dataset, LibriWord, and (2) \sm{AuxSL:} multi-task learning (MTL) with an additional classifier to minimize the domain gap between in-domain and auxiliary data.
Our results have shown that creating a keyword-balanced dataset is a practical approach for training lightweight keyword-spotting models. Moreover, we demonstrated the superiority of the proposed learning approach through extensive experiments, as evidenced by the improved performance in the FS-KWS benchmark.
While our approach has achieved promising results, we recognize that there may be other techniques to mitigate domain differences between datasets, such as RFN~\cite{RFN} and DSBN~\cite{dsbn}. We leave it as future work to explore and analyze the effectiveness of these methods in a scenario where there is a large domain gap between in-domain and auxiliary data.

\bibliographystyle{IEEEtran}
\bibliography{mybib}

\end{document}